**Breaking the superfluid speed limit**

D. I. Bradley, S. N. Fisher, A. M. Guénault, R. P. Haley*, C. R. Lawson, G. R. Pickett, R. Schanen, M. Skyba, V. Tsepelin & D. E. Zmeev

Department of Physics, Lancaster University, Lancaster, LA1 4YB, UK.

* r.haley@lancaster.ac.uk

**Coherent condensates appear as emergent phenomena in many systems[1-8], sharing the characteristic feature of an energy gap separating the lowest excitations from the condensate ground state. This implies that a scattering object, moving through the system with high enough velocity for the excitation spectrum in the scatter frame to become gapless, can create excitations at no energy cost, initiating the breakdown of the condensate[1,9-13]. This limit is the well-known Landau velocity[9]. While, for the neutral Fermionic superfluid $^3$He-B in the *T*=0 limit, flow around an *oscillating* body displays a very clear critical velocity for the onset of dissipation[12,13], here we show that for *uniform linear motion* there is no discontinuity whatsoever in the dissipation as the Landau critical velocity is passed and exceeded. Since the Landau velocity is such a pillar of our understanding of superfluidity, this is a considerable surprise, with implications for the understanding of the dissipative effects of moving objects in all coherent condensate systems.**

The Landau critical velocity marks the minimum velocity at which an object moving through a condensate can generate excitations with zero energy cost[9]. In the frame of the object, moving at velocity $v$ relative to the fluid, excitations of momentum $p$ are shifted, by Galilean transformation, from energy $E$ to ($E - v \cdot p$). Superfluid $^3$He has a BCS dispersion curve[1] with energy minima, $E = \Delta$ at momenta $\pm p_F$. Therefore excitation generation should begin as soon as one energy minimum reaches zero, i.e. when the velocity reaches the Landau critical value, $v_L \approx \Delta/p_F$.

We can investigate $v_L$ in condensates in two limiting regimes, i.e. for the motion of microscopic objects (e.g. ions) or for that of macroscopic objects. For ions, the critical velocity has been observed[10] in superfluid $^4$He at the expected value of $\approx 45$ m s$^{-1}$, and confirmed in superfluid $^3$He-B at 28 bar as consistent with the expected $\approx 71$ mm s$^{-1}$ value[11]. For macroscopic objects, the onset of extra dissipation at $v_L$ in superfluid $^4$He cannot be observed since damping from vorticity becomes prohibitive at much lower velocities. However, while macroscopic objects can be readily accelerated at the lowest temperatures to the much lower critical velocities in superfluid $^3$He, the experimental picture is somewhat misleading.

In superfluid $^3$He, oscillating macroscopic objects do indeed show a sudden increase in damping[12], but at a velocity of only $\approx v_L/3$, arising from the emission of quasiparticle excitations from the pumping of surface excitation driven by the reciprocating motion[13]. Although this mechanism does not involve bulk pair breaking, it has created the impression that a Landau critical velocity has indeed been confirmed in $^3$He, which is not the case.

What should we expect for uniform motion? The textbook prediction suggests that at $v_L$ all details of the process become irrelevant. Condensate breakdown becomes inevitable; the constituent Cooper pairs separate; and the properties rapidly approaching those of the normal liquid. Under our experimental conditions, this should be spectacular, since the damping force in the normal fluid is some five orders of magnitude higher than that of the superfluid. Despite this expectation that at $v_L$ the dissipation should suddenly increase to very high values, here we show that in fact no discontinuity at all is observed in the damping as $v_L$ is exceeded.

The measurements are made in $^3$He-B at zero pressure, between 140 and 190 K (in the quasiparticle ballistic limit) in the cooling cell shown in Fig. 1, see Supplementary Methods.

The macroscopic "moving object" is a wire formed into a rectangular shape. We can move the wire over a range of velocities in two ways: with a single stroke of steady uniform velocity; or by oscillation at 66Hz, the mechanical resonant frequency. For oscillatory motion the damping force as a function of velocity is directly measured from the resonance. For steady motion we infer the damping from the thermal response, as shown in Fig. 2. Since the temperature profile of rise and slow return to equilibrium is invariant in our temperature range, we derive the damping force from the pulse height as explained in Supplementary Methods.

Figure 3 shows the velocity dependence of the damping force on the moving wire for both AC and DC motion at around 150 K. The oscillatory motion shows the expected rapid rise starting at approximately $v_L/3$. However, the DC results are strikingly different, showing only a very slow rise, again starting at a $v \approx v_L/3$, but with no discontinuity whatsoever as $v_L/3$ or $v_L$ are passed.

A hint to the processes involved comes from the oscillatory behaviour, for which we have a reasonable understanding[13]. We illustrate this with a model of the states at the surface of the wire in the region of depressed energy gap. Figure 4 shows the appropriate dispersion curves, for both surface and bulk states, in the frame of the moving wire. For simplicity, we assume that $T = 0$ and that the surface states have zero gap. (Note that owing to the pure potential flow field around a cylinder, when the wire moves at velocity $v$, the liquid at the wire surface moves at a maximum relative velocity of $2v$.)

As the wire moves, the surface-state dispersion curve tips. Elastic collisions with the wire allow excitations to cross the curve (the cross-branch processes of panel b), populating states on the RHS and depleting those on the LHS. Given a constant velocity, at some point the distribution of excitations comes into equilibrium with the wire (panel c'). However, if we accelerate the wire fast enough to prevent these cross-branch processes from maintaining equilibrium, excitations on the LHS can be elevated to energies of $2vp_F$ (panels b and c). Occasional cross-branch processes transfer some of these excitations across to the opposite branch. When the energy of these excitations matches the energy of the RHS dispersion-curve minimum for bulk liquid, which is falling as $\Delta - vp_F$, they can enter the bulk via the escape process (panel c). This loss of local excitations represents dissipation, and occurs as soon as $2vp_F = \Delta - vp_F$, i.e. when $v = \Delta/3p_F$, the "critical velocity" measured for oscillatory motion.

We can draw two conclusions. First, this can only happen if the cross-branch processes are relatively slow, but not too slow. If very fast, the branch distributions would always remain in equilibrium with the wire (as in panel c' where no escape process is possible at $v = \Delta/3p_F$), and if very slow, no branch equalisation occurs at all and again no escape processes are possible.

Secondly, this must be a transient effect, since at constant velocity the cross-branch processes must ultimately prevail and the distribution will become that of panel c'. In other words, as we accelerate the wire we should see a pulse of excitations emitted as soon as the velocity reaches $v = \Delta/3p_F$, but if the velocity increases no further, the number of excitations able to escape will become depleted and the dissipation will cease. Of course, in oscillatory motion this does not happen since on reaching maximum velocity, the motion reverses and the whole process repeats in the opposite direction, with the emission of further excitations.

Now consider the effect of an initial acceleration to a sustained steady velocity. Starting from zero we will see the same behaviour as in panels **a-c** in Fig. 4. As the velocity increases beyond $v_L/3$, surface states over a larger region around the wire can access the escape process, (i.e. not just at the points of maximum surface flow velocity), as in Fig. 5. This increases both the escape probability and the angular range of emission[15], increasing the damping force during acceleration (panels **d** and **e**). When $v_L$ is reached (panel **f**), a new escape process does indeed become available as surface excitations on the LHS can now escape directly into the LHS minimum of the bulk dispersion curve. However, again nothing sudden occurs at this point, only steady growth in the escape probability.

Suppose the acceleration stops to give a final steady velocity above $v_L$. The surface excitation distributions will gradually come into equilibrium with the wire, cutting off the escape processes. Thus in this steady state the dissipation ceases. Subsequently, during the deceleration at the end of the stroke, the converse process comes into play yielding a further burst of escaping excitations.

For finite temperatures we already know the damping force arising from the background of thermally-excited quasiparticles[14] (blue dashed line in Fig. 3). The escape processes add the extra component indicated in Fig. 3, but there is no jump at $v_L$.

We emphasize that these are mechanisms for promoting local surface excitations into the bulk condensate. (This is somewhat akin to the "baryogenesis" analogue seen when excitations localised in vortex cores are ejected when the vortices are moved[16].) Paradoxically, there is no mechanism for breaking of Cooper pairs in the bulk, *despite* the wire moving through the condensate at a velocity above the pair-breaking minimum.

Of course, the Landau argument has to be correct, but it seems that for $^3$He-B, the "boundary layer" of depressed gap shields the bulk superfluid from the ravages of the Landau process. This "shielding" arises because, at near-zero temperature, there is no mechanism for the condensate to gain information about what the moving body is doing on the other side of the boundary layer. Related effects are seen in rotating superfluid $^3$He-B where at very low temperatures the lack of normal fluid disconnects the superfluid from the rotating container[17-19]. This scenario is shown schematically in panel **g** of Fig. 5. Conversely, microscopic objects (smaller than the coherence length) in the liquid have only a marginal disturbing effect on the superfluid order parameter and are thus fully exposed to the bulk condensate.

Our results were consistent up to 190 K. At higher temperatures, the growing normal fluid fraction must come into play to transmit this information to the bulk condensate, allowing the "classical" critical velocity behaviour to emerge. Unfortunately, the greater dissipation from the increasing normal fluid fraction rules out measurements in this regime. Perhaps this effect can only be studied in our "pure" condensate. For the future, the same process could be profitably studied in $^3$He-A where, depending on orientation, the anisotropic order parameter presents either a full BCS gap or nodes where $v_L$ would be zero[20,21].

To conclude, we can definitely state that for a macroscopic object moving in superfluid $^3$He-B approaching the $T = 0$ limit, we see no discontinuity in the damping force when we cross the Landau velocity. This was astonishing to us, as it is such a cornerstone of our understanding of condensates. However, the demonstrated ability to create non-destructive flows beyond the "common knowledge" limit of $v_L$ will lead to new techniques and regimes for investigating superfluids, and will stimulate new studies of coherent condensates in general.

**Acknowledgements** We thank S.M. Holt, A. Stokes and M.G. Ward for excellent technical support. This research is supported by the UK EPSRC and by the European FP7 Programme MICROKELVIN Project, no. 228464.

**Author Contributions** Aided by contributions from all authors, SNF, RPH, CRL and RS devised and built the experiment. AMG, MS, RS and DEZ made the measurements. RPH, MS and RS made the calculations. DIB, AMG, RPH, GRP and VT devised the interpretation and AMG, RPH, and GRP wrote the paper. All authors discussed the results and commented on the manuscript.

**Author Information** Reprints and permissions information is available at www.nature.com/reprints. The authors declare no competing financial interests. S.N. Fisher passed away on 4 January 2015. Readers are welcome to comment on the online version of the paper. Correspondence and requests for materials should be addressed to R.Haley@lancaster.ac.uk.

**Figure 1 | Cutaway of the double demagnetisation cooling cell (outer cell muted).** The adiabatic demagnetisation refrigerant comprises high-purity copper plates coated in sintered silver for thermal contact and immersed in the superfluid $^3$He sample filling the cell and cooled to ≈ 130 K. The moving wire "goalpost" (crossbar width 9 mm) is mounted as shown and is driven perpendicular to its plane through the $^3$He. The temperature rise caused by the dissipation is monitored by the vibrating wire and quartz tuning fork thermometers.

**Figure 2 | The sustained DC velocity method and measurement.** The upper panel shows the very accurate linear stroke of the moving wire, sustained over a distance of 2 mm, with blue bands indicating the periods of initial acceleration and of final deceleration. The lower panel shows the associated temperature response, with the period of the linear motion shown by the red band. The line shape is consistent with a burst of dissipation during the initial acceleration and a second during the final deceleration, see Supplementary Methods.

**Figure 3 | The damping force as a function of velocity for oscillatory (AC) motion compared with steady (DC) motion.** The Landau velocity, $\Delta/p_F$, is marked and the region above the Landau velocity tinted red for clarity. **a**, The measured dissipation for oscillatory (AC) motion shows a very sudden rise when the peak velocity reaches $v_L/3$. The small dissipation arising from the residual thermal gas of quasiparticles is also indicated. Based on the observations for ionic motion in both superfluid $^3$He and $^4$He we might naively expect the dissipation for steady motion to rise rapidly above the Landau velocity, as indicated[11, 12]. **b**, However, our measurements for the steady (DC) motion at a similar temperature show only a slow rise, (arising from the same escape of local excitations which leads to the dissipation in the oscillatory motion, see text) but *no sudden increase* in the damping force, even as the Landau critical velocity of $v_L = \Delta/p_F$ is passed (ringed). We see this behaviour independently of temperature up to the highest velocities and temperatures at which we can measure, i.e. ≈ 2.5 $v_L$ and 190 K.

**Figure 4 | The processes involved in oscillatory (AC) motion.** Dispersion curves for the zero-gap excitations at the cylindrical wire surface and the fully-gapped bulk, in the wire rest frame, using the convention of showing all of the filled states at $T = 0$ (red = quasiparticles, blue = quasiholes). For pure potential flow the maximum fluid velocity is $2v$ when the wire moves at $v$. **a**, wire at rest. **b**, wire moves slowly; surface state curves tip by $2vp_F$ and bulk by $vp_F$; surface state excitations scatter elastically ("Cross-branch"). **c**, excitations start to enter the bulk when $v = \Delta/3p_F$ ("Escape"); dissipation begins. **c'**, quasiparticle distributions reach equilibrium with moving wire; escape processes cease.

**Figure 5 | The processes involved in steady (DC) motion.** Dispersion curves for the zero-gap excitations at the cylindrical wire surface and the fully-gapped bulk, in the wire rest frame, for velocities from $v_L/3$ to greater than $v_L$. **d**, **e**, the number of surface states that can scatter and escape increases. **f**, a new process allows surface states to escape directly into the bulk left-hand branch, but there is no extra pair-breaking in the bulk even though the wire moves faster than $v_L$. **g**, Schematic view of the processes above. In the case illustrated quasiholes near the wire are promoted to higher energies by the tipping of the dispersion curves and, following cross-branch processes, can escape into the bulk. The region of reduced gap shields the bulk condensate from pair breaking by the moving wire.

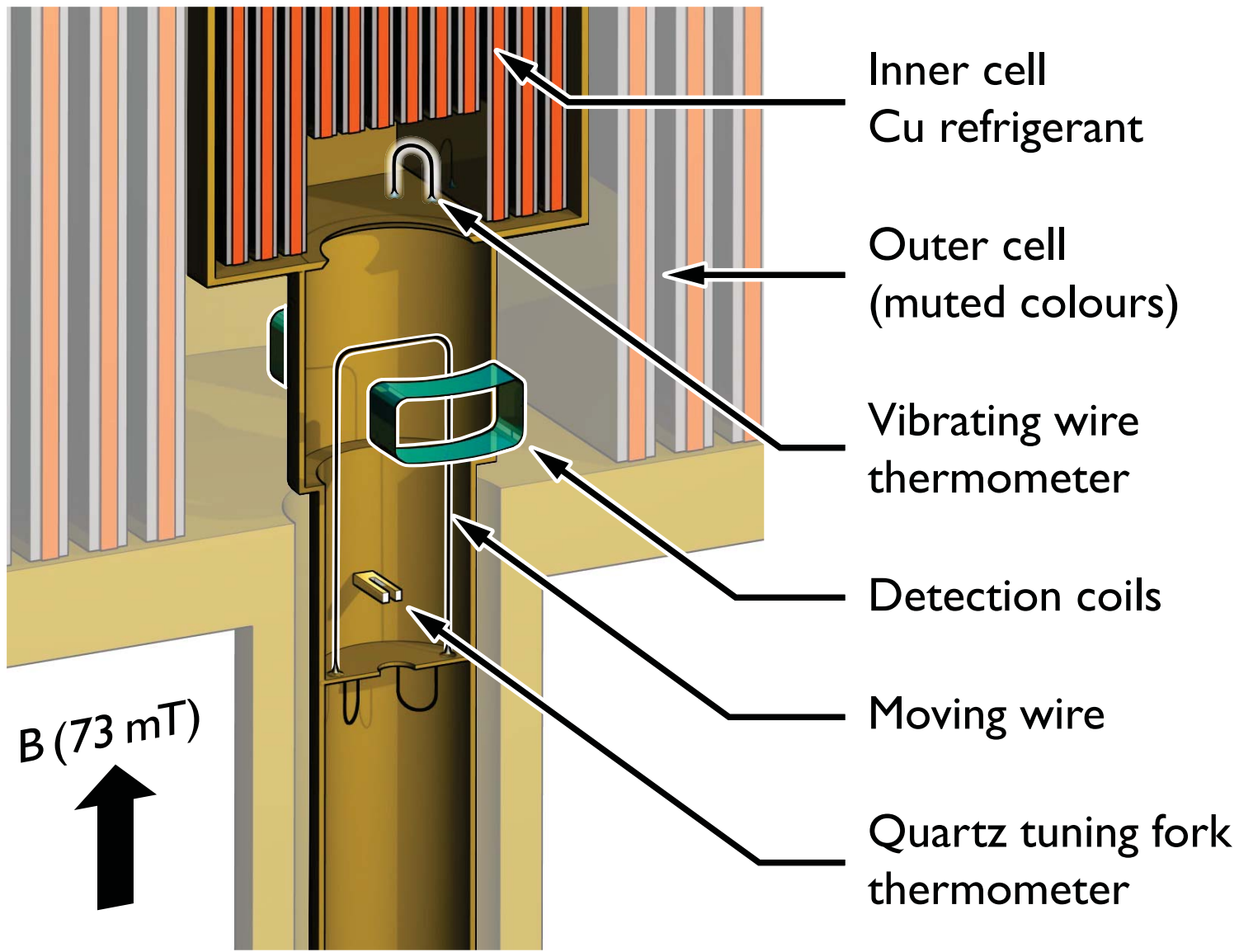

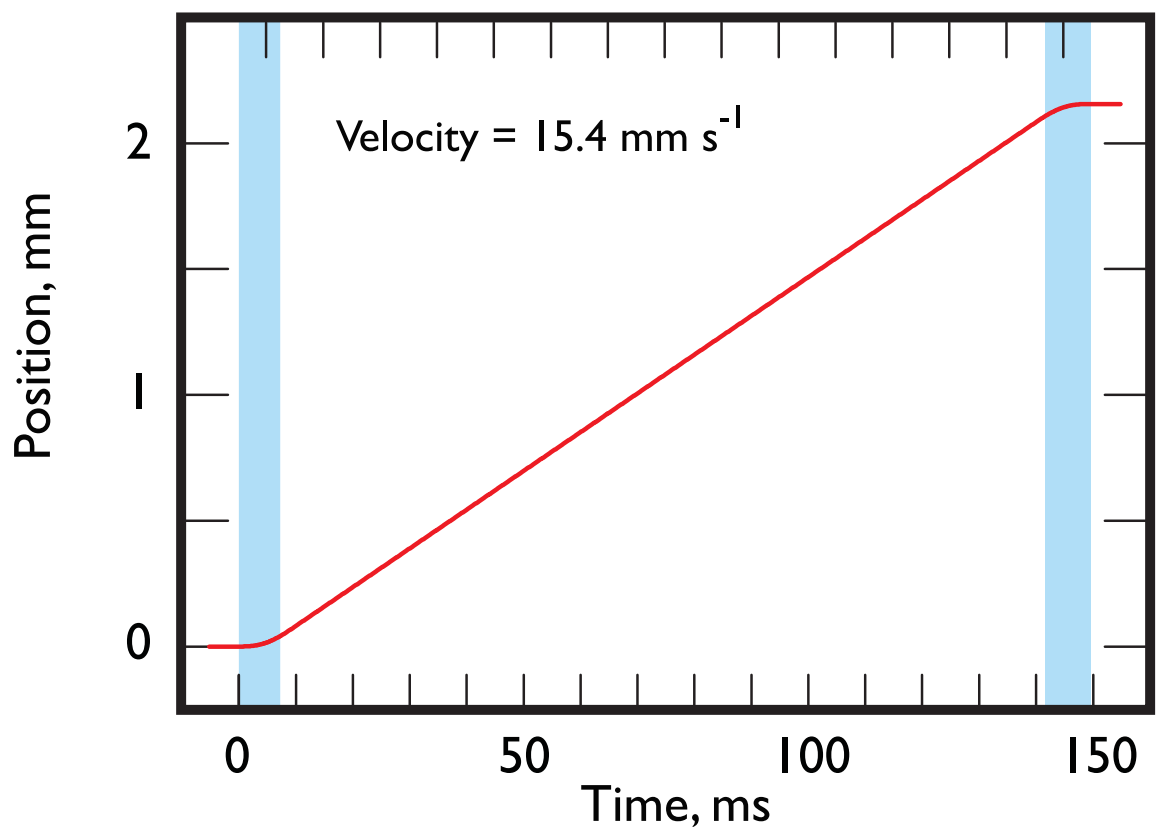

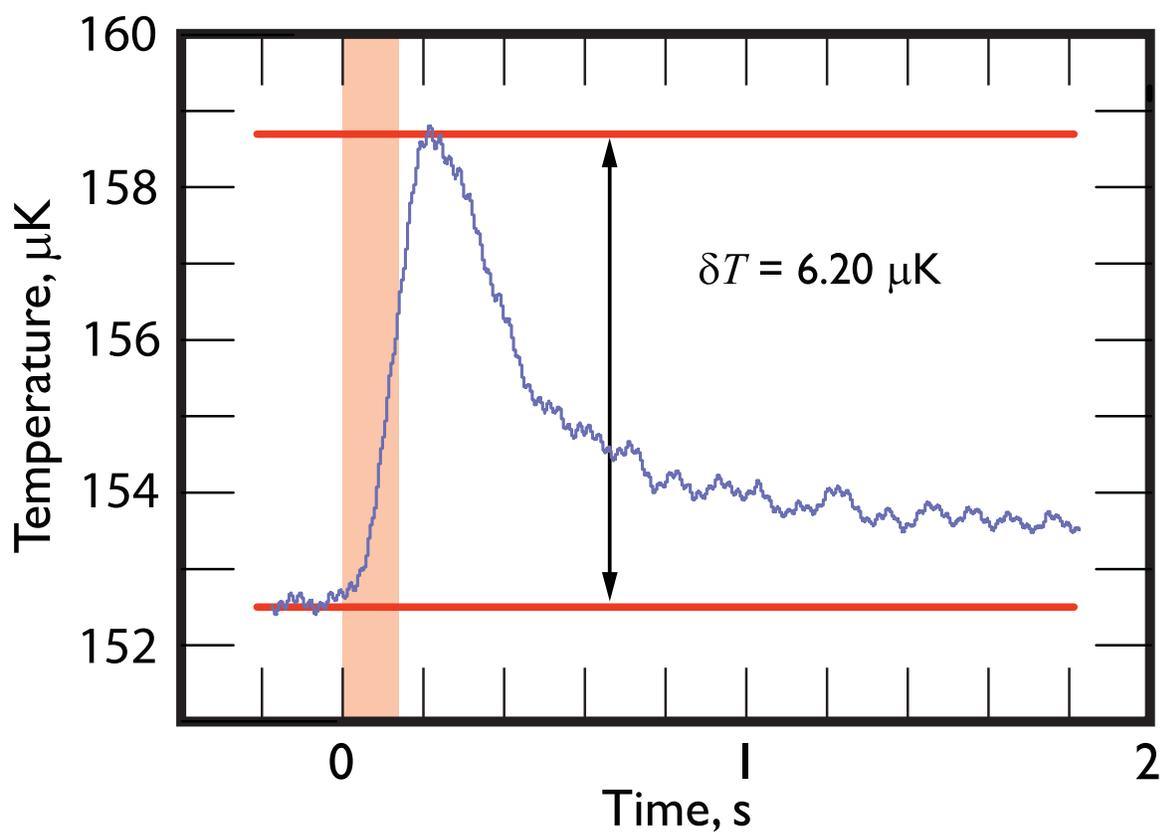

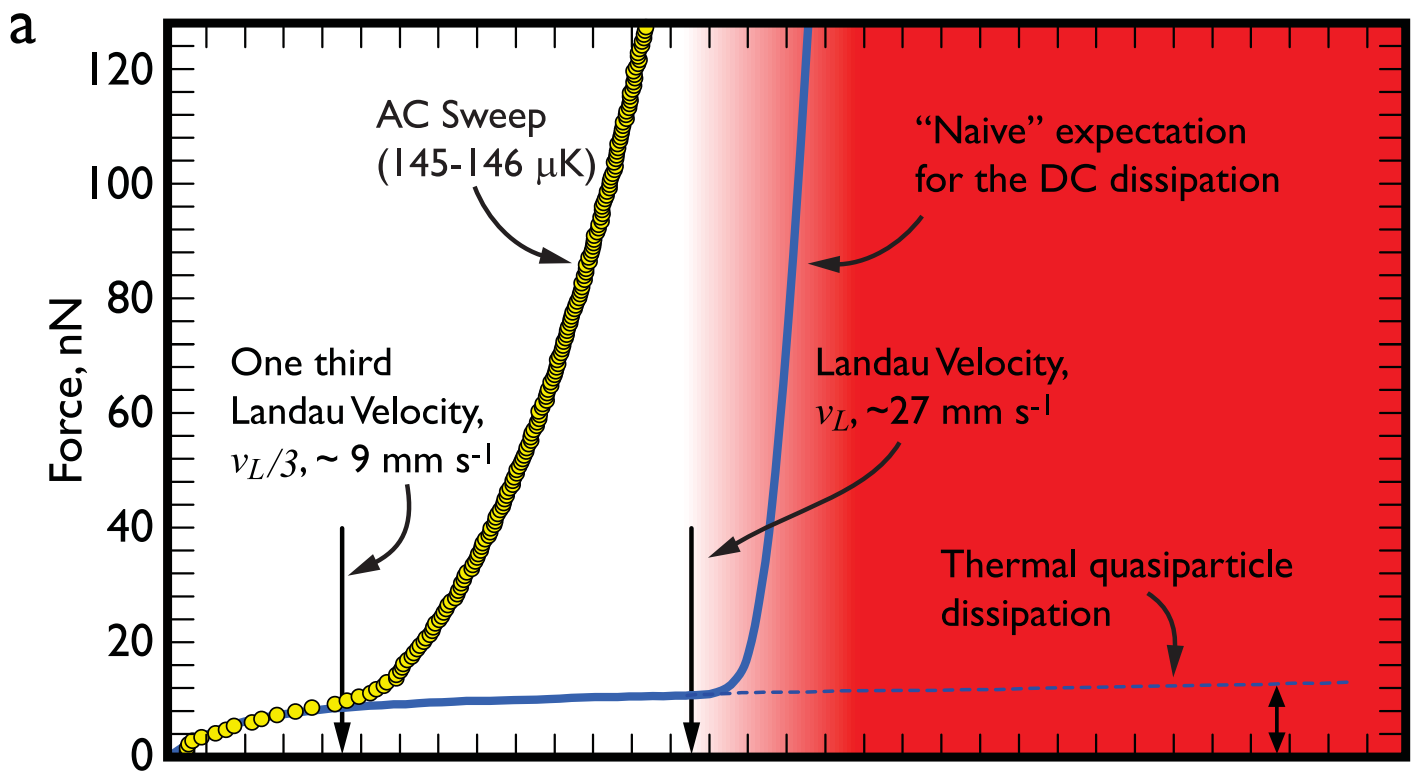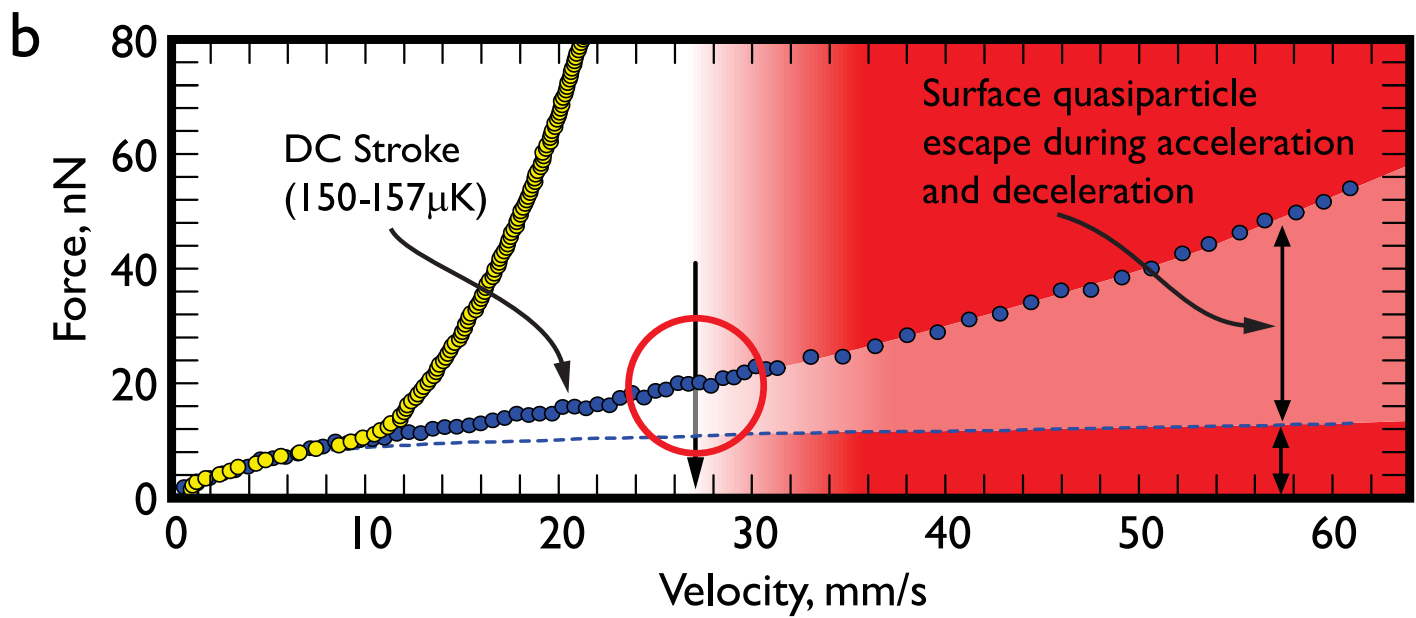

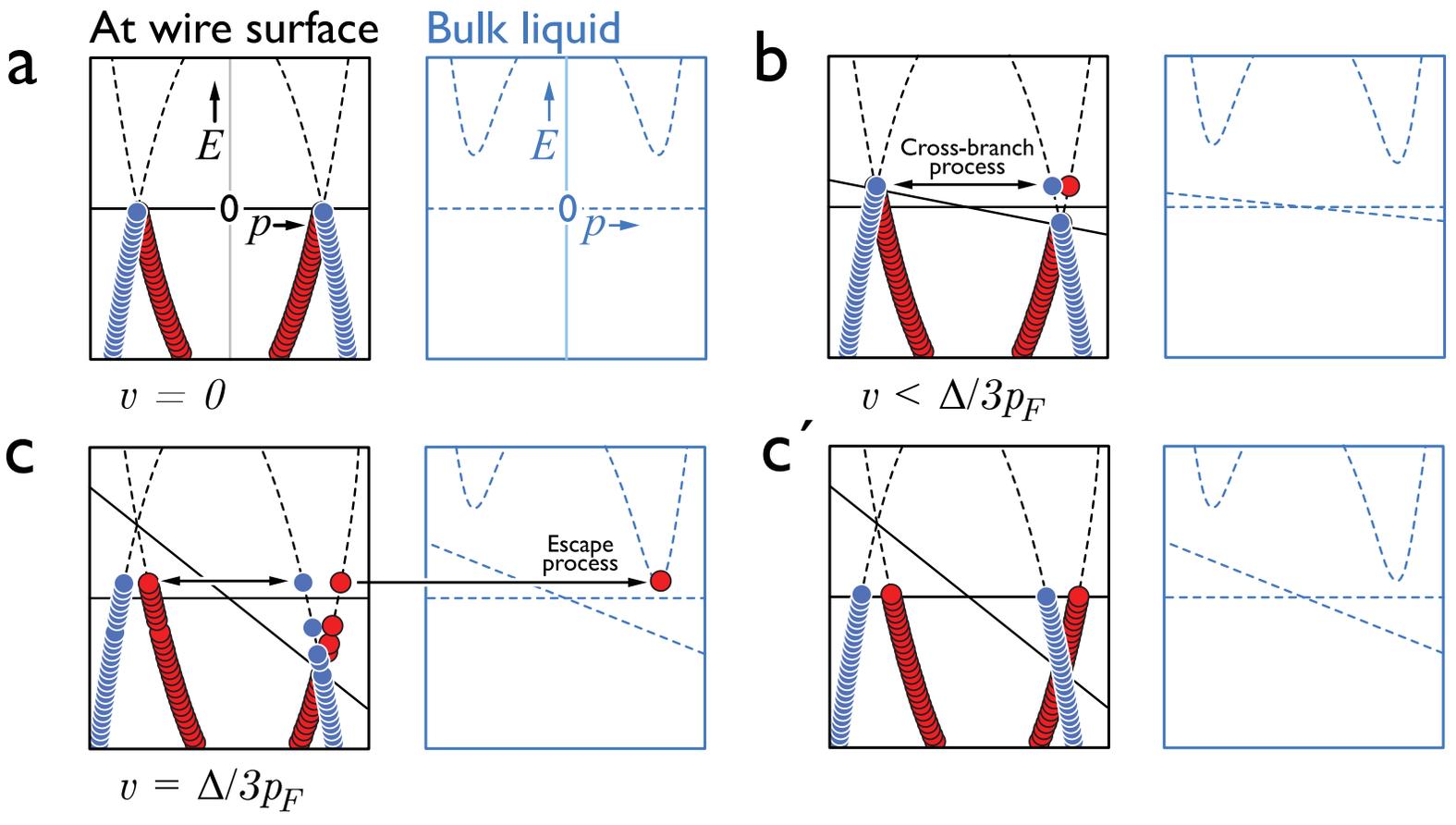

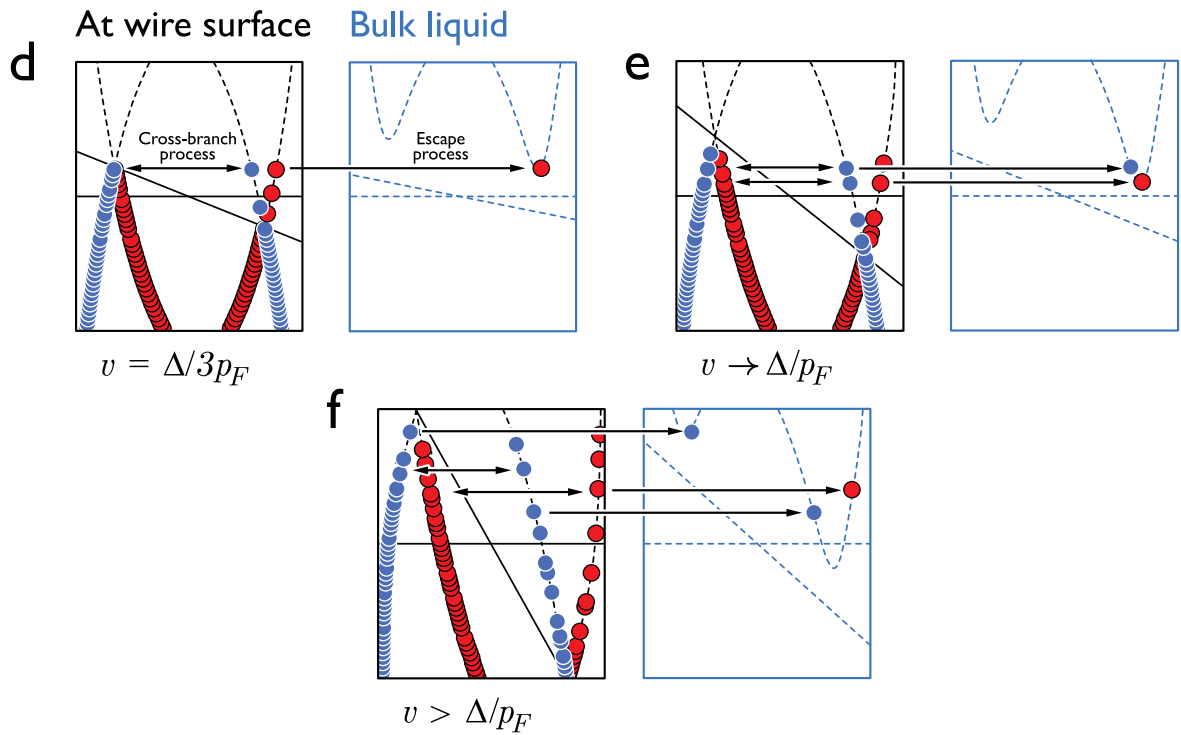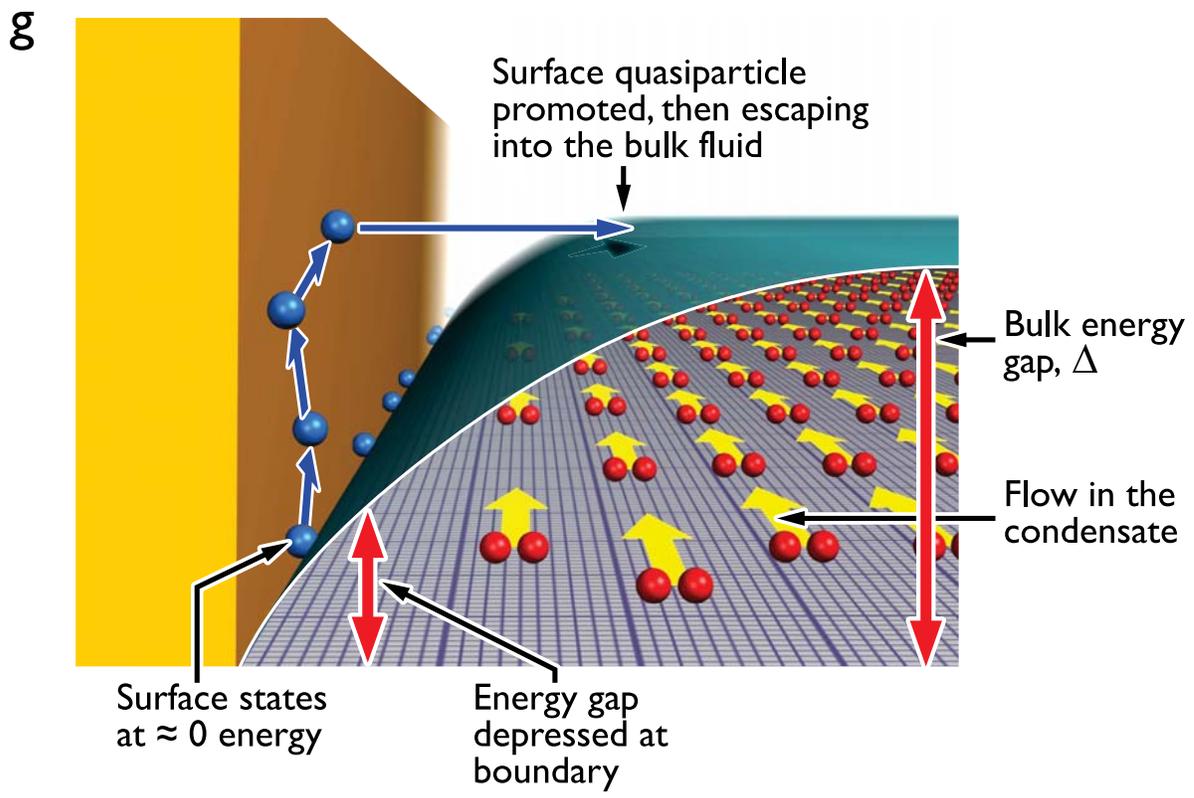

**Supplementary Methods**

The experiments are made in a Lancaster-style nested nuclear demagnetisation cell[S1] for cooling liquid $^3$He, Fig. 1, with the measuring arrangement shown in Fig. S1. The moving wire is mounted inside the lower end of the inner cell. Its position is determined by two detector coils in the outer cell. The wire is a 100 m diameter superconducting NbTi wire bent into a goalpost shape as shown, with height 25 mm and leg spacing 9 mm. In the small vertical external field of 73 mT, a current is applied across the wire and the Lorentz force on the current applies a horizontal force on the wire allowing us to move it or to oscillate it. The experiments are all made in $^3$He-B at 0 bar pressure, in the temperature range of 140 to 190 K, well below the condensate transition temperature $T_c$ = 0.929 mK. This is in the ballistic quasiparticle limit, where the mean free path of excitations is much larger than any of the dimensions of the experimental cell.

For the AC oscillatory mode measurement we use standard techniques[S2,S3] to measure the moving wire at its resonant frequency, 66 Hz. The AC driving current amplitude $I$ is stepped upwards in small steps. The voltage generated across the resonator, as the cross-bar cuts the applied magnetic flux, is determined by a lock-in amplifier at each step. At low velocity, where the wire is very lightly damped, the amplitude sweep must be carried out slowly in order to avoid unwanted ringing. The current is converted to driving force through the relationship $F = BdI$ where $B$ is the magnetic field and $d$ the length of the wire crossbar. The measured AC voltage $V$ is converted to velocity using $v = V/(Bd)$.

For the DC measurements at steady velocity, the position of the wire is inferred from the response of the pick-up coils to a small high frequency (92 kHz) signal added to the driving current. To calibrate the wire position, we gradually increase the drive current until the wire touches the cell wall, at which point the signal in the pick-up coil increases no further, a clear signature that the wall has been reached. We then repeat the process in the opposite direction. Knowing the two extreme positions, we can now accurately determine the position of the wire crossbar at any intermediate point, and we can thereby also derive the spring constant of the wire.

To make a measurement, we move the wire at a constant velocity from a starting point to an end point a few mm away. The ramp provides a short acceleration period, followed by a period of constant velocity, ending with a short deceleration stage, as shown in Fig. 2. Accelerating and maintaining the constant velocity requires a carefully profiled current ramp, computed from the dynamical properties of the wire-fluid system. This active control removes transient effects at the beginning and end of the ramp to ensure that the wire never moves faster that the target DC velocity. This scheme was devised by RS, and a similar scenario is described by DEZ[S4]. Following each such stroke, the current is slowly ramped back to the starting point and paused for a few minutes to allow the cell to return to thermal equilibrium before the next measurement. We log the output data from the driving current, and that from the high frequency lock-in amplifier which follows the rapid changes of the wire position.

The quantity we wish to measure is the dissipation generated by the linear motion. This we track from the resonant response of a nearby 4.5 m NbTi filament vibrating wire resonator. This acts as a "thermometer" (or quasiparticle density detector)[S2,S3] in the superfluid, providing the quantitative measure of the thermal disturbance caused by each stroke. We can also use the quartz tuning fork[S5] shown in Fig. S1 with similar results.

From the thermal dissipation produced we can infer the effective damping force on the wire for each stroke. This conversion requires one calibration constant which we determine by comparison with the damping force for the oscillatory motion (which is measured directly) at the same temperature. We know the quasiparticle damping very accurately at low velocities ($v \ll v_L$)[S6] and thus we scale the damping force to agree with that measured for the oscillatory motion at low velocities (say, as for the AC sweep and DC stroke data from Fig. 3 which are at similar temperatures).

Although we only see what appears to be a single thermal transient, our picture of the process implies that we should see an initial burst of dissipation during acceleration and a second burst during deceleration. Unfortunately the thermometer time-constant time is just too long to resolve such a two pulse shape. However, we have confirmed that the overall shape of the measured pulse is consistent with a convolution of two similar-shaped pulses at the beginning and end of the stroke.

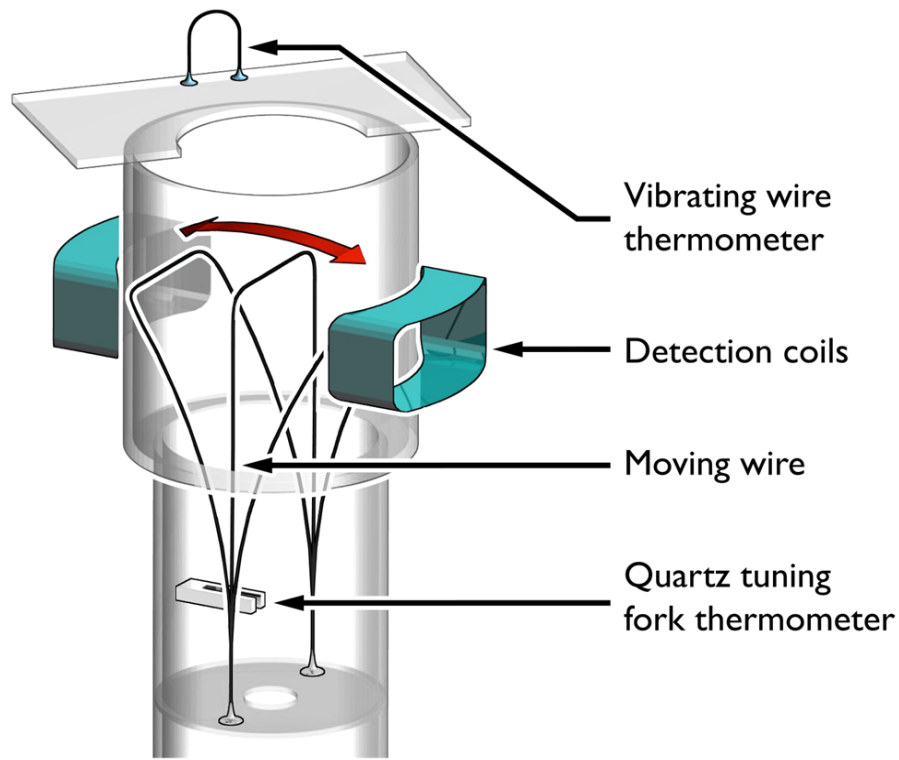

**Figure S1 | A view of the moving wire configuration.** The moving wire is mounted in the inner cell and the crossbar can move over a horizontal distance of ±6 mm before touching the inner cell wall. Its position is sensed by detection coils in the outer cell. The temperature, or thermal quasiparticle density, is monitored by vibrating wire and quartz tuning fork resonators. We calibrate the position of the wire by moving it until it is stopped by the cell wall in each direction (see text) which provides the two fixed points needed to determine the absolute position.